\documentclass[letter]{aa} 
\pdfoutput=1

\usepackage{graphicx}
\usepackage{txfonts}

\begin{document} 

\title{SDSS J1001+5027: Strong microlensing-induced chromatic variation caught in the act}    
                                                             
\author{ Luis J. Goicoechea\inst{1,2} 
   	     \and  Vyacheslav N. Shalyapin\inst{1,2,3}
       }

\institute{
	Instituto de F\'isica de Cantabria (CSIC-UC), Avda. de Los Castros s/n, 
	E-39005 Santander, Spain\\
	\email{goicol@unican.es;vshal@ukr.net}
	\and
	Departamento de F\'\i sica Moderna, Universidad de Cantabria, 
	Avda. de Los Castros s/n, E-39005 Santander, Spain
	\and
	O.Ya. Usikov Institute for Radiophysics and Electronics, National 
	Academy of Sciences of Ukraine, 12 Acad. Proscury St., UA-61085 
	Kharkiv, Ukraine
          }
                
\titlerunning{SDSS J1001+5027: strong microlensing event}
\authorrunning{Goicoechea \& Shalyapin}


\abstract{We conducted long-term monitoring of the doubly imaged gravitationally 
lensed quasar SDSS J1001+5027 consisting of spectro-photometric observations 
separated by $\sim$120 days (time delay between both quasar images), as well as 
test and auxiliary data. This monitoring approach allowed us to reliably find a 
strong microlensing-induced chromatic variation of the quasar continuum in the 
period 2022$-$2025. The ongoing microlensing event has caused the delay-corrected 
spectral flux ratio in 2025 to have a dramatic changing look, opening the door to 
very promising observations of the system in the coming years. These future 
follow-up observations of such a rare event are expected to provide critical 
information to discuss, among other things, the structure of the inner accretion 
flow towards the central supermassive black hole in SDSS J1001+5027.}
   

\maketitle

\section{Introduction}
\label{sec:intro}

The gravitational field of a foreground galaxy can split the emission of a 
background quasar into several images. This effect is called strong gravitational 
lensing and has been the subject of intensive study in the last 45 years 
\citep[e.g.][]{2024SSRv..220...87S}. Because the multiple images originated 
from the same quasar, one might expect their spectra to be identical. 
However, there are two main reasons for differences between image spectra. First, 
the light from the quasar takes different times to reach the observer along 
different paths, and at a given observation time we see image spectra 
corresponding to the quasar emission at different times. The quasar emission is 
time-dependent and this intrinsic variability produces spectral differences 
between images. Second, the quasar images pass through different regions of the 
lensing galaxy, so dust extinction and lens magnification are expected to depend 
on the image considered. Although the macrolens magnification of each image 
(related to the overall distribution of mass in the galaxy) is constant, its 
microlensing magnification (associated with the local population of compact 
objects, e.g. stars) may vary with time and wavelength 
\citep[e.g.][]{2024SSRv..220...14V}.

Sometimes intrinsic and microlensing variabilities modify the spectra of a quasar 
image in a similar way. Intrinsic variations generate changes in the continuum 
slope, and the image usually becomes bluer as it gets brighter 
\citep[e.g.][]{2004ApJ...601..692V}. Additionally, microlensing effects depend on 
the emitting source size and may magnify the inner regions of the accretion disc 
more than the outer ones, also causing a bluer-when-brighter chromatism 
\citep{1991AJ....102..864W}. Due to intrinsic and/or microlensing effects, the 
emission lines may also vary but to a lesser extent than the continuum 
\citep[e.g.][]{2012A&A...544A..62S}. Fortunately, there is a standard method of 
disentangling the intrinsic quasar variability from microlensing variations in a 
doubly imaged gravitationally lensed quasar. The key idea is to observe the 
double quasar at epochs separated by the time delay between its two images, A and 
B, and thus obtain AB spectra at the same emission time 
\citep[e.g.][]{1991AJ....101..813S}. For each pair of time-delay-separated 
epochs, one can divide the spectrum of the trailing image at the second epoch by 
the spectrum of the leading image at the first epoch, effectively removing the 
intrinsic flux of the quasar, and building a spectral flux ratio that only 
accounts for a constant macrolens flux ratio, non-variable chromatic extinction 
effects, and microlensing effects.

The realisation of the above key idea is not easy in practice. Modern telescopes 
have overloaded observing schedules, and organising long-term spectroscopic 
observations with a large telescope at fixed time separations is not trivial. In 
addition, there are flux calibration issues. Spectra taken at different epochs may 
have calibration biases that cannot be removed by calculating ratios between them. 
We note that \citet{2008A&A...480..647E,2008A&A...490..933E} conducted long-term 
spectro-photometric monitoring of the four images of the lensed quasar 
\object{Q2237+0305} (Einstein Cross) using the Very Large Telescope (VLT) of the 
European Southern Observatory (ESO). However, for this system, there are no 
observing time constraints and significant calibration problems because time 
delays are less than one day and simultaneous VLT/ESO spectra of the quasar images 
were used to derive spectral flux ratios.

We solved the telescope scheduling problem by taking advantage of the flexible 
real-time schedule of the 2.0 m Liverpool Telescope (LT) operating in robotic mode 
\citep{2004SPIE.5489..679S}. We also developed a dedicated flux calibration 
pipeline to decrease calibration errors of LT spectra. To our knowledge, no 
previous observing programme has been carried out to systematically obtain 
double-quasar spectra separated by a standard time delay of several weeks or 
months. Our LT long-term spectro-photometric monitoring programme focused on the 
three widely separated, bright double quasars \object{Q0957+561}, \object{SDSS 
J1001+5027}, and \object{SDSS J1515+1511} with time delays of 4$-$14 months 
\citep{2018A&A...616A.118G}. This paper presents the timely and compelling 
results for \object{SDSS J1001+5027}. 

\section{SDSS J1001+5027}
\label{sec:object}

The doubly imaged gravitationally lensed quasar \object{SDSS J1001+5027} was 
discovered by \citet{2005ApJ...622..106O} and has been intensively observed over 
the last 20 years. The two quasar images (A and B) are separated by 2\farcs928 
\citep{2016MNRAS.458....2R}, while the redshifts of the source quasar and the main 
lensing galaxy are $z_{\rm s}$ = 1.841 and  $z_{\rm l}$ = 0.415, respectively 
\citep{2012AJ....143..119I}. Optical light curves of A and B were also used to 
accurately measure a time delay of $\Delta t$ = 119.3 $\pm$ 3.3 d \citep[A is the 
leading image;][]{2013A&A...557A..44R}.

Regarding early optical observations, the spectra of both images included 
\ion{C}{iv}, \ion{C}{iii}], and \ion{Mg}{ii} emission lines, but the brightest 
image, A, showed a more pronounced continuum towards blue than the dimmer image, B 
\citep[see the discovery spectra in Fig. 2 of][]{2005ApJ...622..106O}. 
\cite{2005ApJ...622..106O} also estimated the B/A spectral flux ratio at a single 
epoch, which gradually increased from 0.5 at 4000 \AA\ up to 0.9 at 8000 \AA, with 
a slight excess at $\sim$4400 \AA\ (\ion{C}{iv} emission). This overall behaviour 
of the spectral flux ratio suggested that image B could be affected by dust in the 
lensing galaxy. More recently, \citet{2018A&A...616A.118G} reported evidence of
extinction of the B-image continuum during the first 10 years of observations of 
the double quasar. In addition, \object{SDSS J1001+5027} is a broad absorption 
line (BAL) quasar, and \cite{2017MNRAS.468.4539M} and \cite{2018ApJ...854...69M} 
studied its absorption features in detail and found variable \ion{C}{iv} outflow 
lines.

\section{Spectro-photometric monitoring}
\label{sec:monitor}

\begin{figure}
\centering
\includegraphics[width=9cm]{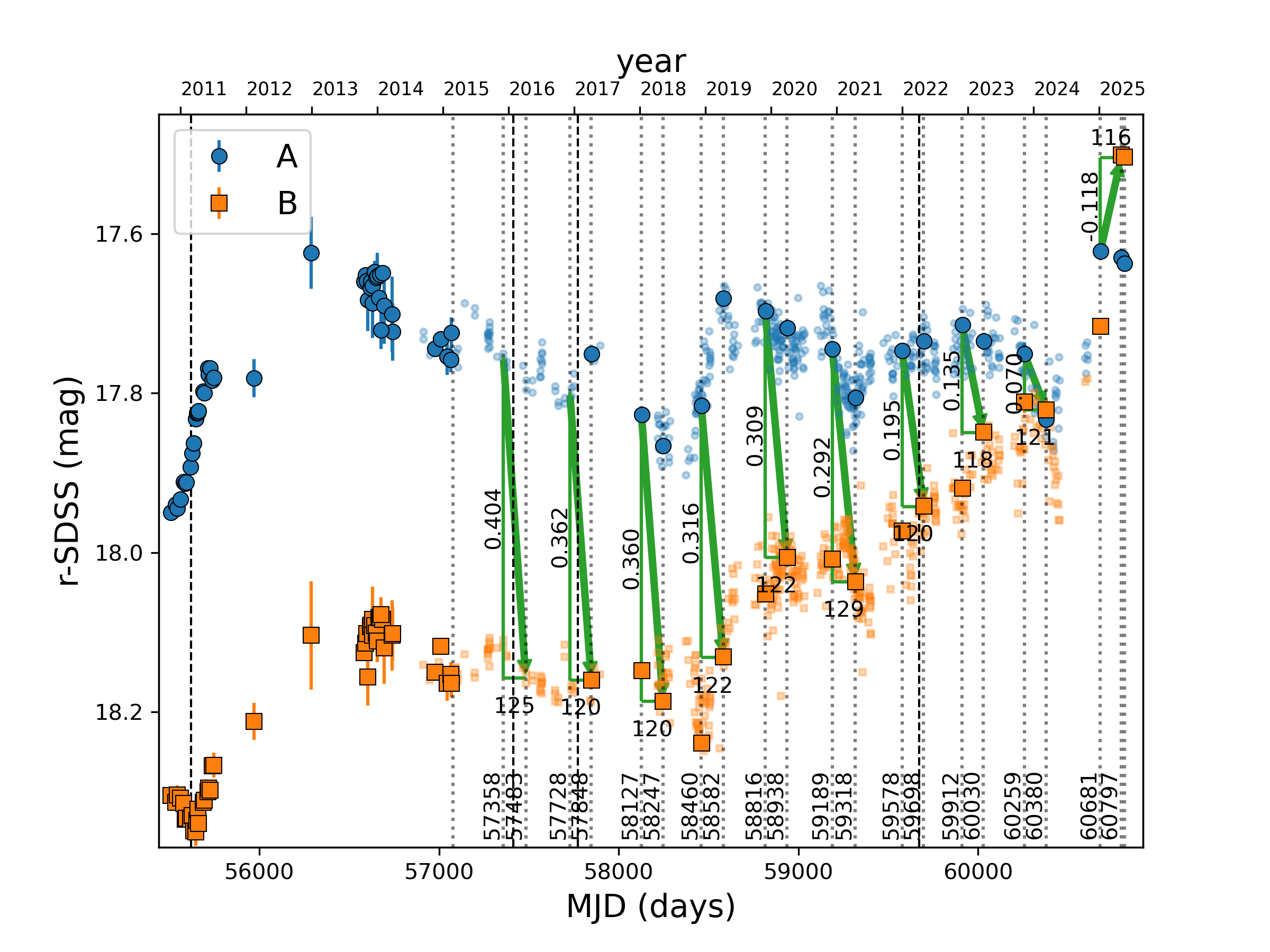}
\caption{Light curves of \object{SDSS J1001+5027} in the $r$ band and 29 
spectroscopic epochs. The large circles/squares with black borders represent 
LT-HCT-PS magnitudes, while the small semi-transparent circles and squares 
describe Gaia-ZTF magnitudes (see main text for details). The vertical dotted 
lines show 25 epochs of spectroscopic observations with LT/SPRAT, leading to ten 
pairs of AB spectra separated by the time delay of the system (green arrows and 
horizontal green bars). For these AB pairs, the $r$-band magnitude offsets are 
highlighted with vertical green bars. The dashed vertical lines show four epochs 
of auxiliary spectroscopy (see Appendix~\ref{sec:appendix2}).}
\label{fig:lcs}
\end{figure}

In the framework of the Gravitational LENses and DArk MAtter (GLENDAMA) project 
\citep{2018A&A...616A.118G}, the first spectro-photometric observations of 
\object{SDSS J1001+5027} were made with the LT in the period 2013$-$2014, using 
the FRODOSpec integral-field unit and the IO:O CCD camera. 
Unfortunately, the spectra extracted from our 3000-s FRODOSpec exposures turned 
out to be noisy and of limited astrophysical interest. The situation improved 
significantly with the installation of the SPRAT long-slit spectrograph 
\citep{2014SPIE.9147E..8HP} and since 2015 we have mostly used SPRAT in blue mode 
to regularly perform pairs of observations separated by about 4 months (the time 
delay of the system). This resulted in a total of ten pairs over the last ten 
observing seasons (see the green arrows in Figure~\ref{fig:lcs}). In 
Figure~\ref{fig:lcs}, we display all SPRAT observation epochs (vertical dotted 
lines), including five test epochs. 

We usually took 4$\times$600 s spectroscopic exposures at each observation epoch, 
covering a wavelength range from 3950 to 8100 \AA\ with a resolving power of 350 
($\Delta \lambda \sim 18$ \AA\ at the centre of the spectrum). The slit width and 
position angle were 1\farcs8 (4 pixel) and 122\degr (or 302\degr), so it was 
oriented along the line joining both quasar images. A Xe arc lamp exposure and a 
standard star were used for wavelength calibration and instrument response 
correction, respectively. The spectroscopic observations were generally 
accompanied by 2$\times$250 s photometric exposures with IO:O in the SDSS $r$ 
band. These IO:O exposures allowed us to extract quasar image magnitudes 
\citep[e.g.][]{2018A&A...616A.118G} that were used to calibrate the SPRAT spectra 
and follow up on the $r$-band variability of the two quasar images. Our 
spectroscopic data reduction is fully described in Appendix~\ref{sec:appendix1}.

In Figure~\ref{fig:lcs}, the large circles and squares with black borders show all 
LT/IO:O magnitudes between 2013 and 2025, along with magnitudes from the 2.0 m 
Himalayan $Chandra$ Telescope (HCT) in 2010$-$2011 \citep{2013A&A...557A..44R} and 
the Pan-STARRS \citep[PS;][]{2020ApJS..251....7F} 
database\footnote{http://panstarrs.stsci.edu} at four epochs. 
Gaia\footnote{https://gea.esac.esa.int/archive/} \citep{2016A&A...595A...1G} and 
the $Zwicky$ Transient 
Facility\footnote{https://irsa.ipac.caltech.edu/cgi-bin/Gator/nph-dd} 
\citep[ZTF;][]{2019PASP..131a8003M} brightness records in 2014$-$2017 and 
2018$-$2024, respectively, are also plotted in Figure~\ref{fig:lcs} using small 
semi-transparent circles and squares. Red magnitudes from telescopes other than 
the LT were shifted to the LT $r$-band system. For the most relevant 
spectro-photometric data of A and B (see the green arrows in 
Figure~\ref{fig:lcs}), the time separations between A and B are highlighted with 
horizontal green bars, whereas the magnitude offsets between A and B are 
highlighted with vertical green bars.

\section{Pairs of AB spectra separated by the time delay}
\label{sec:spectra}

\begin{figure}
\centering
\includegraphics[width=9cm]{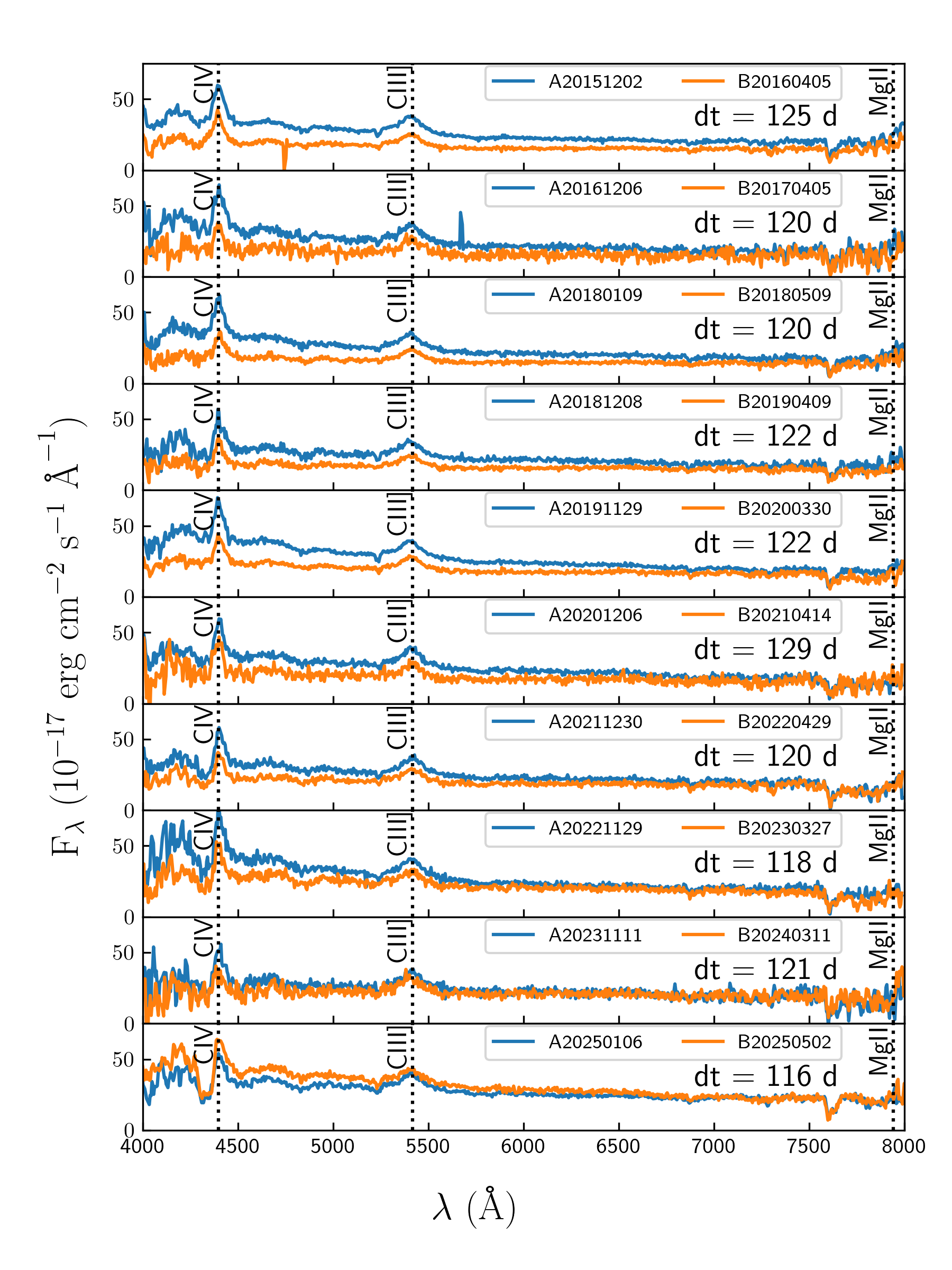}
\caption{Pairs of time-delay-separated AB spectra. Each pair consists of the 
spectrum of the leading image A at a first epoch and the spectrum of the trailing 
image B at a second epoch separated by $dt$ from the first.}
\label{fig:abspectra}
\end{figure}

We obtained ten pairs of AB spectra separated by $dt \sim \Delta t$ (see the 
horizontal green bars in Figure~\ref{fig:lcs})\footnote{Quasar images barely vary 
(if at all) at optical wavelengths on a time scale of $\sim$5 d}, which are 
displayed in Figure~\ref{fig:abspectra}. Each AB pair corresponds to image spectra 
at the same emission time, so differences between them are due to differential 
astrophysical effects. Both differential dust extinction and 
differential macrolens magnification are expected to be constant over time, while 
differential microlensing effects may vary with time. Thus, chromatic microlensing 
variability of \object{SDSS J1001+5027} is caught in action in 
Figure~\ref{fig:abspectra}. The spectral differences between A and B have 
undergone a clear chromatic evolution over the last 10 years, starting with image 
B being fainter and redder than image A, and ending with image B being brighter 
and bluer than image A.  

\subsection{Emission lines}
\label{sec:lines}

The SPRAT wavelength range completely covers only the \ion{C}{iv}$\lambda$1549 and 
\ion{C}{iii}]$\lambda$1909 emissions. These two emission lines have complex 
profiles. The blue wing of the \ion{C}{iv} line is significantly affected by 
absorption features in the BAL quasar \citep[e.g.][]{2018ApJ...854...69M}, and 
the \ion{C}{iii}] line profile has two blue-wing excesses related to 
\ion{Si}{iii}]$\lambda$1892 and \ion{Al}{iii}$\lambda$1857 emissions. Thus, we 
focused on the cores of the \ion{C}{iv} and \ion{C}{iii}] emission lines. The core 
of each emission line was defined as the spectral region within the rest-frame 
velocity interval of $[-1000, +1000]$ km s$^{-1}$, so the line core flux comes 
mainly from a large region unaffected by microlensing. This corresponds to the
narrow line region and the outer parts of the broad line region.

\begin{figure}
\centering
\includegraphics[width=9cm]{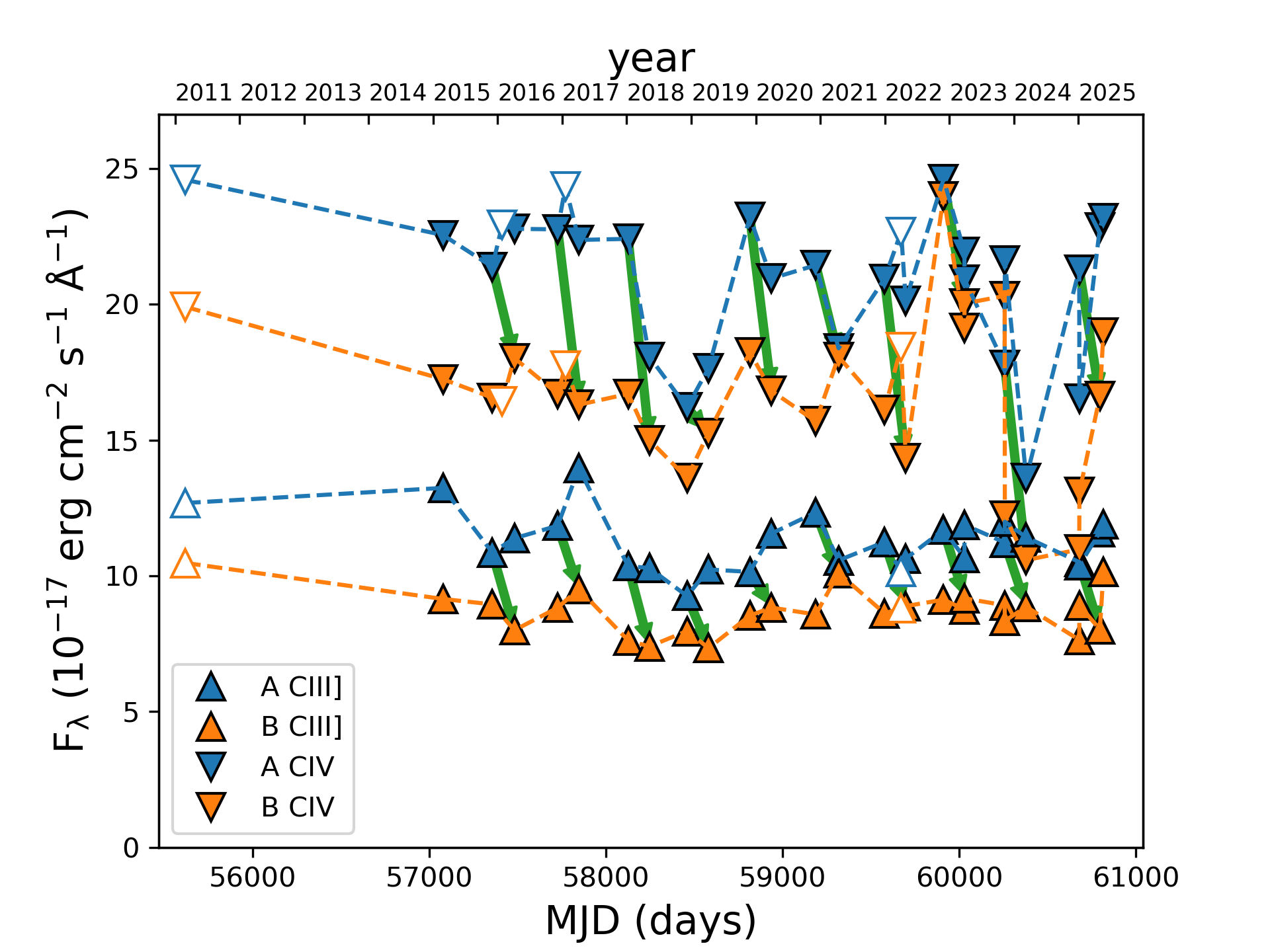}
\caption{Fluxes of carbon emission line cores. In addition to the measurements 
from the LT/SPRAT spectra (filled triangles), fluxes associated with the auxiliary 
spectra are shown (open triangles; see Appendix~\ref{sec:appendix2}).}
\label{fig:carbon}
\end{figure}

\begin{figure}
\centering
\includegraphics[width=9cm]{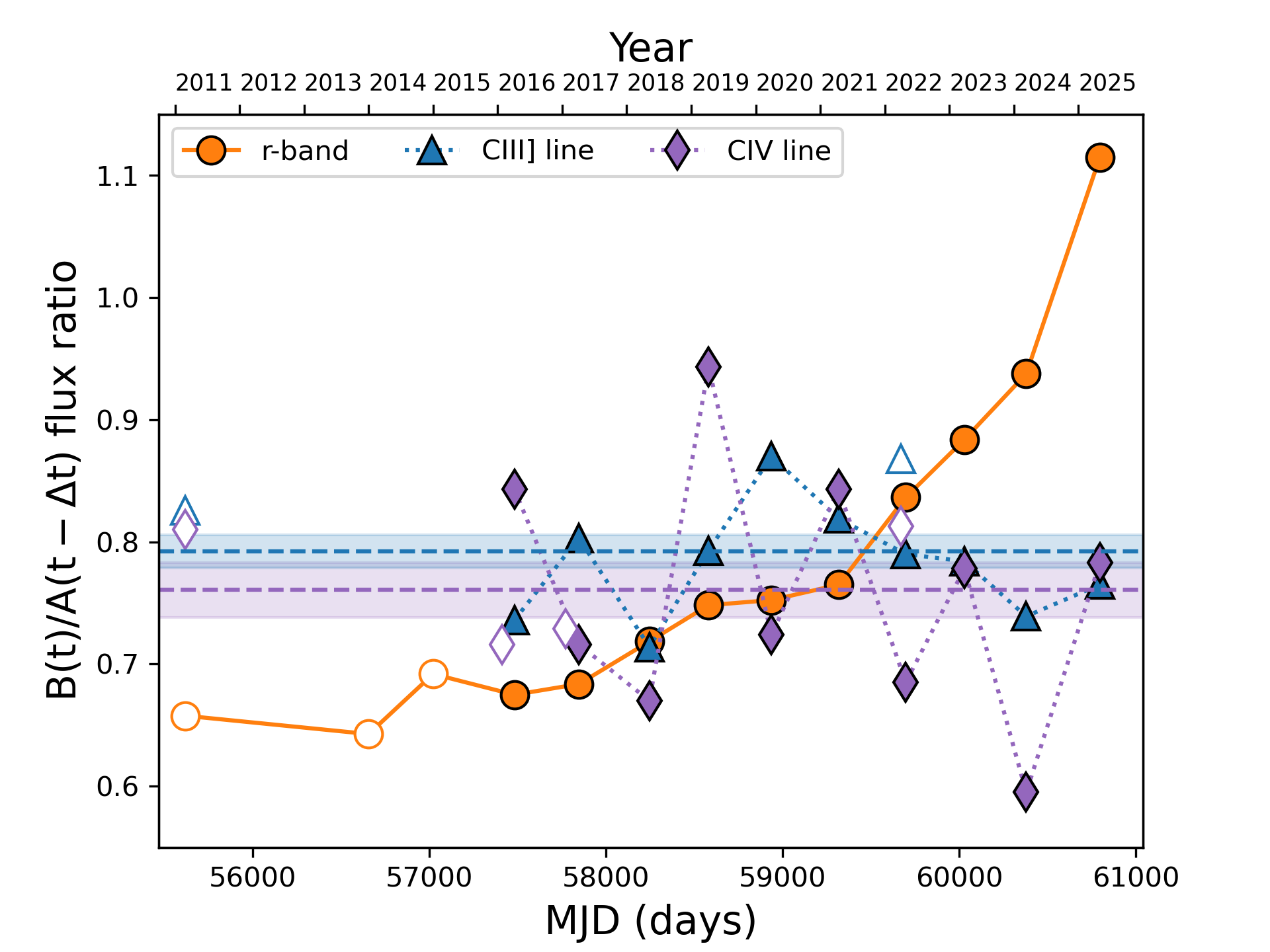}
\caption{Flux ratios for the cores of the \ion{C}{iv} and \ion{C}{iii}] emission 
lines. The filled purple diamonds and filled blue triangles represent 
delay-corrected \ion{C}{iv} and \ion{C}{iii}] line-core flux ratios, respectively, 
and the horizontal dashed lines and strips describe the average values and 
their uncertainties (purple for \ion{C}{iv} and blue for \ion{C}{iii}]). The open 
purple diamonds (\ion{C}{iv}) and open blue triangles (\ion{C}{iii}]) 
represent single-epoch measurements from auxiliary spectra (see 
Appendix~\ref{sec:appendix2}). For comparison, we also show the long-term 
evolution of the delay-corrected $r$-band flux ratio (filled circles), along with 
a few single-epoch values before 2016 (open circles).}
\label{fig:carboncont}
\end{figure}

The line core flux for each carbon ion is shown in Figure~\ref{fig:carbon}. The 
large scatter in the \ion{C}{iv} flux of the two quasar images most probably does 
not have a physical origin but is due to large calibration uncertainties near the 
blue edge of the spectrograph (see Appendix~\ref{sec:appendix1}). In addition, the 
delay-corrected B/A flux ratios for the \ion{C}{iv} and \ion{C}{iii}] line cores 
(filled purple diamonds and filled blue triangles in Figure~\ref{fig:carboncont}) 
show considerable dispersions 
around their average values (horizontal dashed purple and blue lines in 
Figure~\ref{fig:carboncont}). Since there is no apparent long-term variability, we 
assumed that the observed dispersions are dominated by calibration noise, and 
estimated delay-corrected line-core flux ratios of 0.761 $\pm$ 0.023 (\ion{C}{iv}) 
and 0.792 $\pm$ 0.014 (\ion{C}{iii}]). These two estimates can be interpreted as 
the product between the macrolens flux ratio and the extinction ratio at the 
central wavelength of each emission line. However, the two measures are consistent 
with each other and with the single-epoch flux ratio at near-IR wavelengths 
\citep[0.787 $\pm$ 0.05 in the $K$ band at 21\,200 \AA;][]{2016MNRAS.458....2R}, 
suggesting that dust extinction is not appreciably affecting large regions 
containing carbon ions and $\sim$0.79 is a good proxy of the macrolens flux ratio.

\subsection{Continuum}
\label{sec:cont}

\begin{figure}
\centering
\includegraphics[width=9cm]{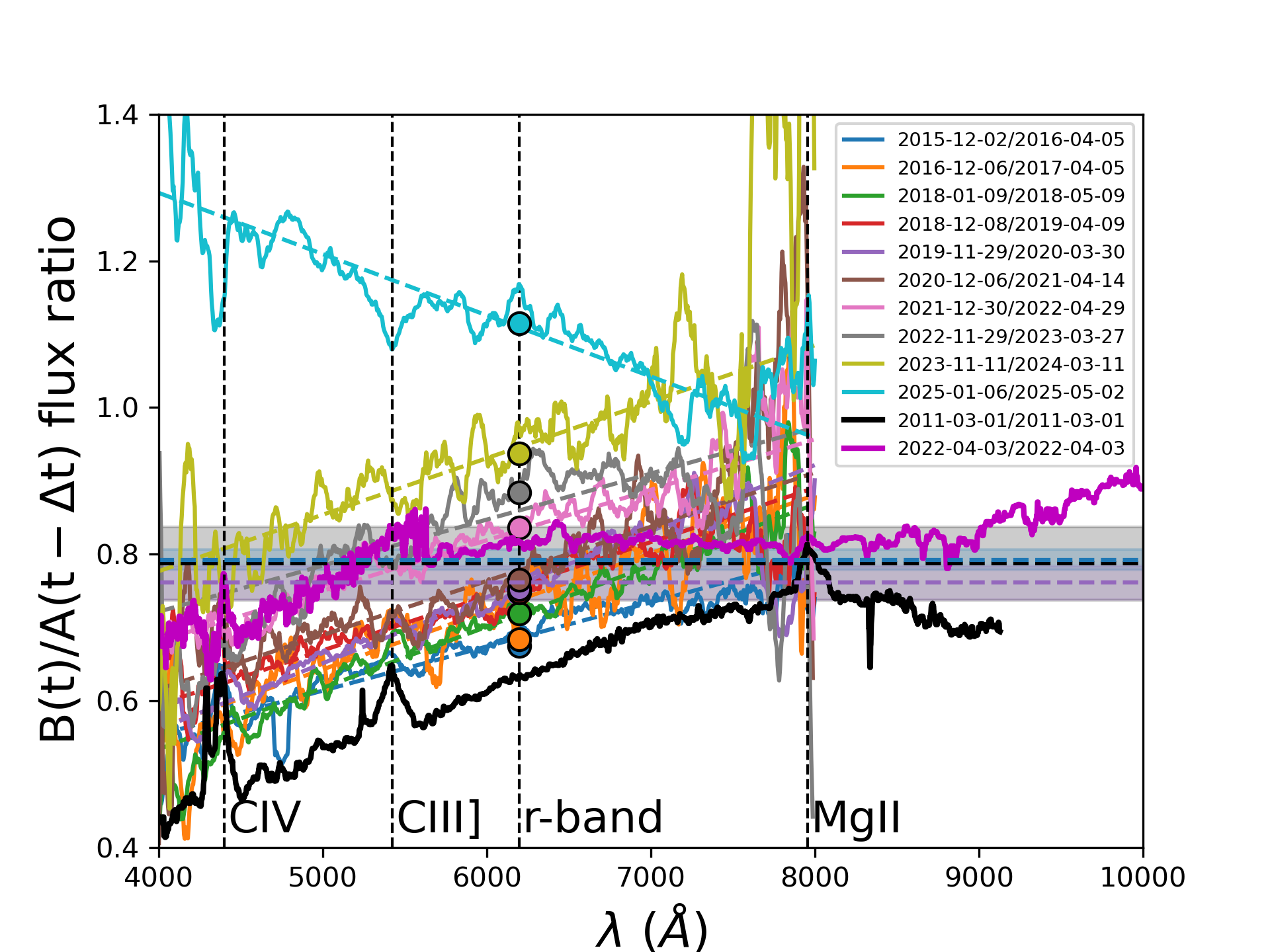}
\caption{Delay-corrected spectral flux ratio of \object{SDSS J1001+5027}. The 
delay-corrected $r$-band flux ratio (filled circles), and the single-epoch 
spectral flux ratio from Gemini North/GMOS (black line) and Keck/LRIS (deep purple 
line) are also shown. The three horizontal strips correspond to the 
\ion{C}{iv} (purple), \ion{C}{iii}] (blue), and $K$-band (grey) flux ratios (see 
the end of Section~\ref{sec:lines}), and are used to highlight the macrolens flux 
ratio.}
\label{fig:specfluxrat}
\end{figure}

The delay-corrected spectral flux ratio is depicted in 
Figure~\ref{fig:specfluxrat}. In this figure, for comparison purposes, we also 
show the single-epoch spectral flux ratio from auxiliary spectra on 1 March 2011 
(Gemini North/GMOS; black line) and 3 April 2022 (Keck/LRIS; deep purple line). 
These Gemini North-Keck data are presented in Appendix~\ref{sec:appendix2}. 
Figures~\ref{fig:carboncont} and \ref{fig:specfluxrat} also include the 
delay-corrected $r$-band flux ratio (filled circles). This broadband ratio is 
primarily associated with continuum emission and has increased monotonically over 
the past nine years. Additionally, the continuum slope in the delay-corrected 
spectral flux ratio has also changed over that time period, undergoing a drastic 
change in 2025 (cyan line at the top of Figure~\ref{fig:specfluxrat}). 

To understand the physics behind the observed evolution (changes) of the 
continuum, which is assumed to mainly come from the accretion disc, we considered 
our proxy of the macrolens flux ratio ($\sim$0.79; see the horizontal strips in 
Figure~\ref{fig:specfluxrat}). At the beginning of the monitoring period, the B 
image was de-magnified (relative to A) by dust extinction effects; it then 
underwent microlensing-induced changes over the last nine years. The 
delay-corrected flux ratio at $\sim$6200 \AA\ ($r$ band) increased noticeably in 
the period 2022$-$2025, while it increased considerably more at the 
shortest wavelengths and barely changed at the longest ones (see 
Figure~\ref{fig:specfluxrat}). The blue portion of the spectrum of image B arises 
from inner rings of the accretion disc and was magnified more by microlensing than 
the red portion arising from more external rings. This recent chromatic 
microlensing episode led to a changing-look of the (delay-corrected) spectral flux 
ratio. At present, image B is brighter and bluer than A, just the opposite of what 
was observed in time-delay-separated AB spectra in the first monitoring years (see 
Figure~\ref{fig:abspectra}). 

\section{Conclusions and discussion}
\label{sec:end}

One of the main aims of the GLENDAMA project was to detect microlensing-induced 
quasar chromatic variablity through the analysis of pairs of optical spectroscopic 
observations of a double quasar separated by the time delay between its two images 
\citep{2018A&A...616A.118G}. Here we have presented a long-term spectro-photometric 
follow-up of \object{SDSS J1001+5027} with the LT, leading to flux ratios between 
quasar images that are not affected by quasar intrinsic variability. The 
delay-corrected $r$-band flux ratio has increased significantly over recent years 
and the continuum slope in the delay-corrected spectral flux ratio has changed 
dramatically in the last year of monitoring. These two results suggest that the 
source quasar is passing through a region of high microlensing magnification 
(e.g. crossing a microlensing caustic). 

Although microlensing-induced chromatic variations in lensed quasars were 
theoretically predicted long ago \citep{1991AJ....102..864W}, they have so far 
only been confidently identified in one quadruply imaged quasar \citep[the 
Einstein Cross; e.g.][]{2008A&A...490..933E}. In this lens system, the lensing 
galaxy is located at a very low redshift and the cadence of strong microlensing 
variations is relatively high. In fact, multi-band light curves of the Einstein 
Cross over a period of 14 years provided a detailed description of several 
microlensing chromatic variations and compelling evidence for a double 
caustic-crossing event \citep{2020A&A...637A..89G}. In this paper, we performed 
the complicated task of disentangling intrinsic variability, dust extinction, and 
microlensing effects \cite[e.g.][]{2008A&A...478...95Y} in the doubly imaged 
quasar \object{SDSS J1001+5027}, unambiguously detecting microlensing chromatic 
variability. Detailed monitoring of this ongoing microlensing event with X-ray, 
UV, optical, and IR facilities offers us a unique opportunity to improve our 
understanding of the structure of quasars \citep[e.g.][]{2024SSRv..220...14V}.

\begin{acknowledgements}
We thank the referee for drawing our attention to several details. We also thank 
the staff of the Liverpool Telescope (LT) for a kind interaction. The LT is 
operated on the island of La Palma by Liverpool John Moores University in the 
Spanish Observatorio del Roque de los Muchachos of the Instituto de Astrof\'isica 
de Canarias with financial support from the UK Science and Technology Facilities 
Council. This long-term research has been supported by the University of Cantabria
and several research grants, including the grant PID2020-118990GB-I00 funded 
by MCIN/AEI/10.13039/501100011033.
\end{acknowledgements}

\begin{appendix}

\section{Spectoscopic data reduction}
\label{sec:appendix1}

The LT data reduction pipeline removed low-level instrumental signatures through 
bias and dark current subtraction, and flat-fielding. Additionally, we cleaned 
cosmic rays using a Python version of L.A.Cosmic algorithm 
\citep{2001PASP..113.1420V} called 
Astro-SCRAPPY\footnote{https://doi.org/10.5281/zenodo.1482019} and carried out 
other standard calibration procedures in a NOIRLab IRAF v2.18 environment 
\citep{2024arXiv240101982F}. The outputs are wavelength-calibrated sky-subtracted 
2D spectra.

\subsection{Source spectra extraction}
\label{sec:specext}

After the initial spectral reduction (see above), the 1D spectrum of the standard
star was extracted by summing fluxes along the slit for each wavelength bin. 
However, the extraction of 1D spectra of both quasar images is a more complex task 
\citep[e.g.][]{2014A&A...568A.116S}. For the double quasar, the light 
distribution along the slit was modelled as two Gaussian-shaped profiles with the 
same width and separated by 2\farcs928 (see Sect.~\ref{sec:object}). This spatial 
model depends on four parameters: the Gaussian width, the two Gaussian amplitudes, 
and the position of the centre of the Gaussian profile for the image A. In a first 
iteration, we fitted the four free parameters of the spatial model to the fluxes 
along the slit for each wavelength bin. The Gaussian width and the position of A 
were then approximated by low-degree polynomial functions of the observed 
wavelength, and in a second iteration, these two parameters were fixed to their 
polynomial values, fitting only the two Gaussian amplitudes. The spatial 
integration of the Gaussian profiles at each wavelength bin led to the 1D spectra 
of A and B.  

\subsection{Flux calibration issues}
\label{sec:caliss}

We paid special attention to the instrument response functions from observations 
of the standard star and the corrections of chromatic slit losses caused by 
differential atmospheric refraction (DAR). These calibration issues are described 
here below.

\begin{figure}
\centering
\includegraphics[width=9cm]{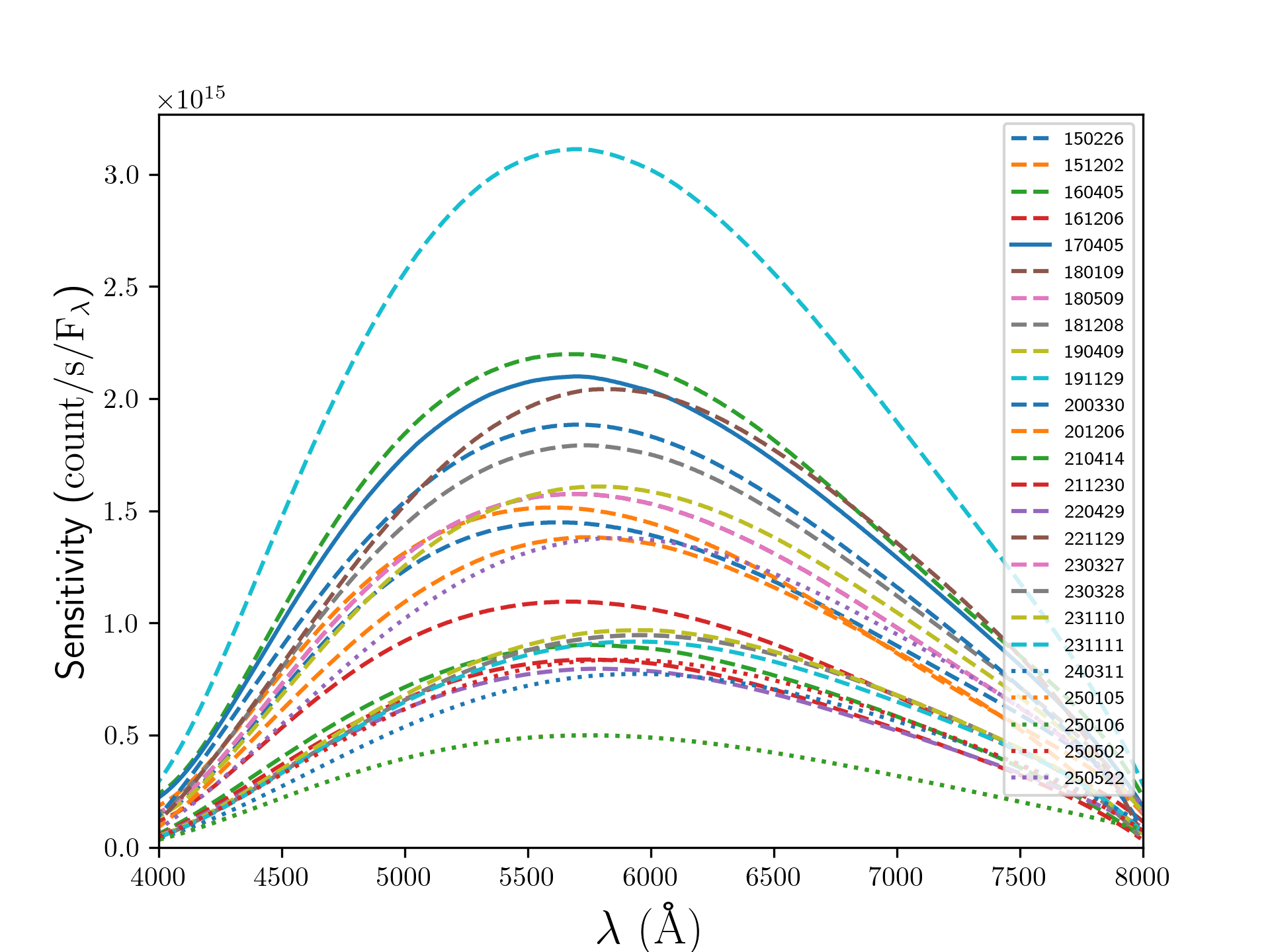}
\caption{SPRAT response functions at the 25 observation epochs. We use labels 
YYMMDD, where YY, MM, and DD represent the last two digits of the year, the month, 
and the day of the month, respectively. These YYMMDD labels are also used in 
Figures~\ref{fig:rsens}, \ref{fig:offset}, and \ref{fig:slitloss}.}
\label{fig:sens}
\end{figure}

\begin{figure}
\centering
\includegraphics[width=9cm]{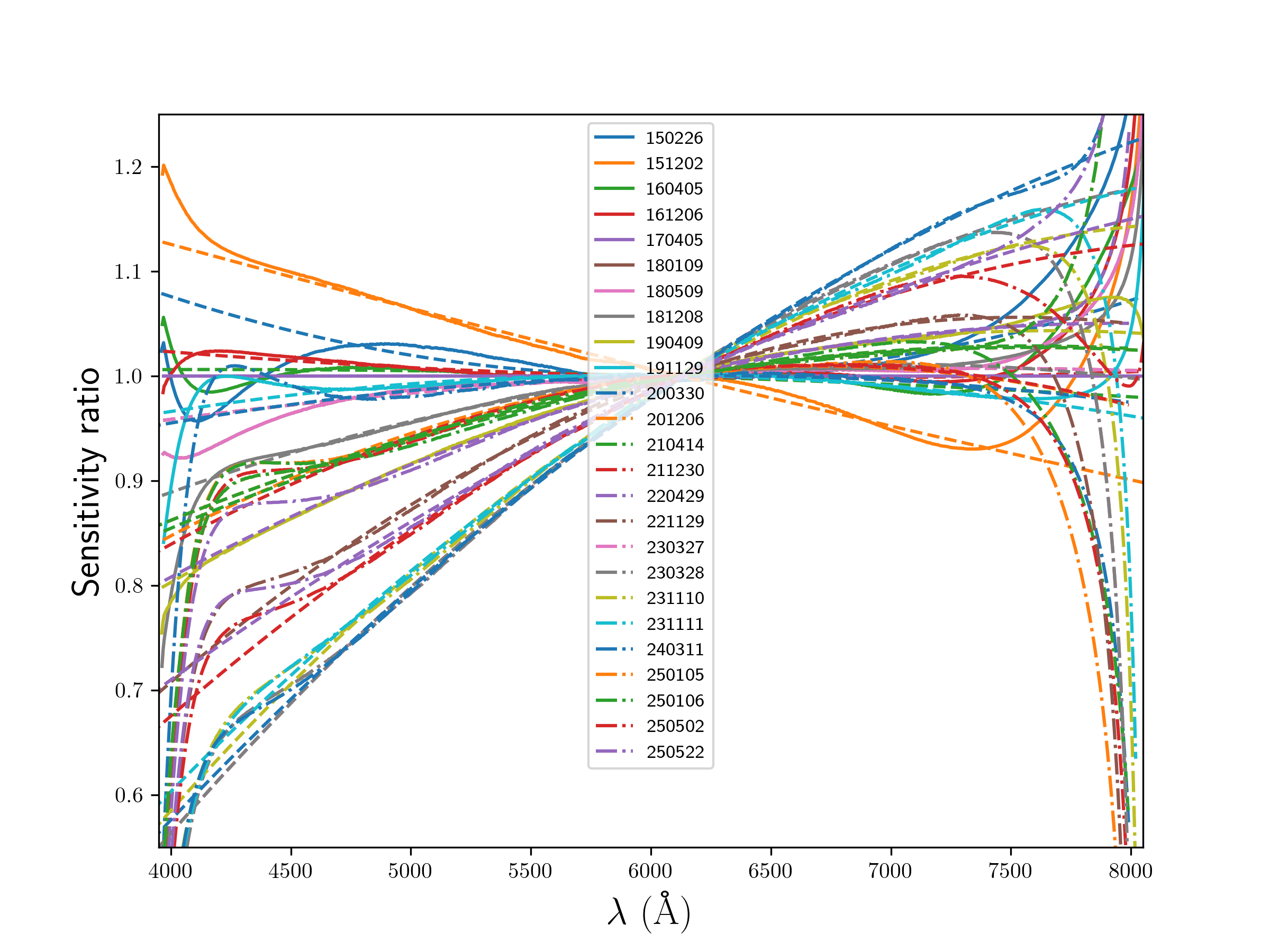}
\caption{Normalised relative responses (see main text). The dashed lines are 
second-degree polynomial fits to the normalised relative responses in the central 
spectral region (4500$-$7500 \AA), which are extrapolated to both spectral edges.}
\label{fig:rsens}
\end{figure}

The SPRAT response function on a given night was initially estimated from standard 
star data for that night, and the response functions at all SPRAT observation 
epochs are shown in Figure~\ref{fig:sens}. However, the extraction of the 1D 
stellar spectra is not reliable at the low-sensitivity spectral edges, probably 
leading to biases in the instrument responses at the shortest and longest 
wavelengths. To analyse possible biases, we took as reference the 170405 response 
function (corresponding to good observing conditions on 5 April 2017; see the blue 
solid line in Figure~\ref{fig:sens}), and then calculated the relative responses 
(dividing by the reference) and multiplied them by appropriate factors so that 
they equal 1 at 6200 \AA. These normalised relative responses are plotted in 
Figure~\ref{fig:rsens}. Due to the evidence of instrumental artefacts at 
4000$-$4500 and 7500$-$8000 \AA, we fitted second degree polynomials to the 
normalised relative responses in the central spectral region (4500$-$7500 \AA) and 
extrapolated these central fits to the spectral edges (dashed lines in 
Figure~\ref{fig:rsens}). Hence, for each night, the normalised response function 
was finally obtained as the product of the reference response and the 
corresponding polynomial function. 

\begin{figure}
\centering
\includegraphics[width=7cm]{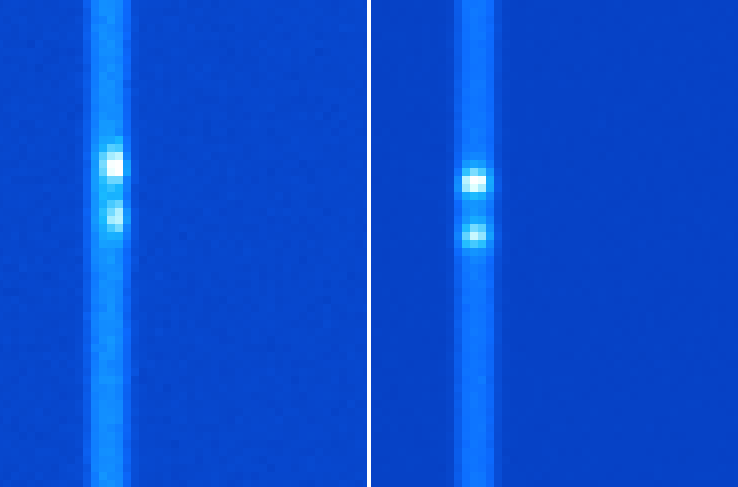}
\caption{Through-slit acquisition frames on 9 April 2019 (left) and 2 December 
2015 (right). In the left panel, the two quasar images are clearly off-axis.}
\label{fig:slit}
\end{figure}

The LT is a robotic telescope, and it is difficult to place sources on the slit 
axis by an automatic unattended operation. Thus, the quasar images are expected 
not to be perfectly centred on the slit. To estimate their position across the 
slit at each observation epoch, we analysed the corresponding through-slit 
acquisition frame (see Figure~\ref{fig:slit}). First, we payed attention to the 
transverse brightness distribution of the background light far from the quasar and 
found the position of the slit axis on the SPRAT CCD ($y$ = 0). After subtracting 
the background, we then determined the acquisition transverse offset $\delta y$ of 
the quasar images by fitting an off-axis Gussian profile to the light distribution 
across the slit in the region where the quasar images are located. The two quasar 
images are assumed to be shifted by $\delta y$ at a reference wavelength of 6200 
\AA, which is close to that of maximum sensitivity of the SPRAT CCD 
\citep{2014SPIE.9147E..8HP} and to the central wavelength of the $r$ passband (the 
final flux calibration is based on photometric measurements with IO:O in the $r$ 
band; see below).

\begin{figure}
\centering
\includegraphics[width=9cm]{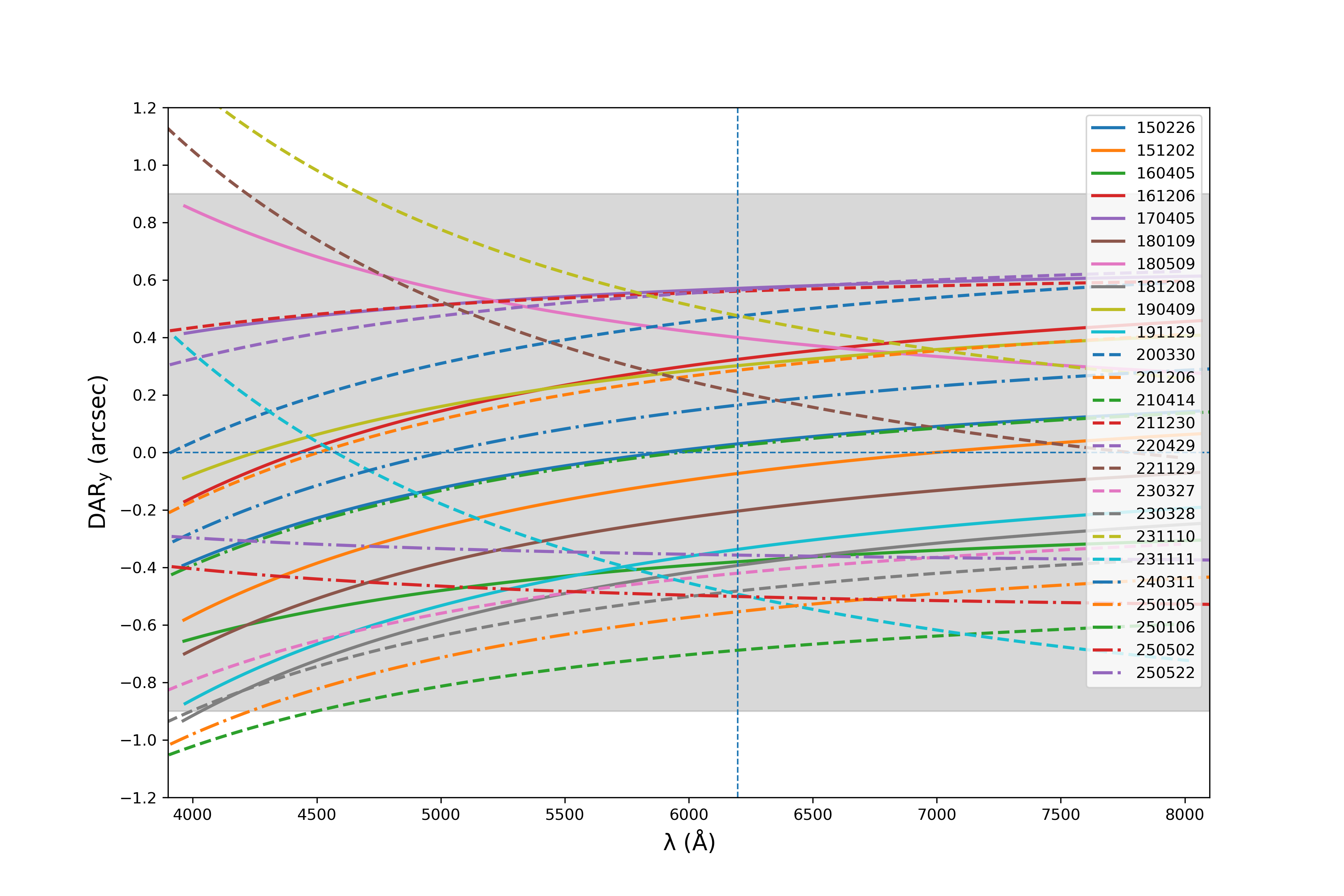}
\caption{Chromatic offsets of quasar images across the slit for each observation 
night. These offsets are due to off-axis acquisitions at 6200 \AA\ and DAR. The 
horizontal dashed line represents the slit axis, while the vertical side of the 
grey strip corresponds to the slit width of 1\farcs8.}
\label{fig:offset}
\end{figure} 

\begin{figure}[h!]
\centering
\includegraphics[width=9cm]{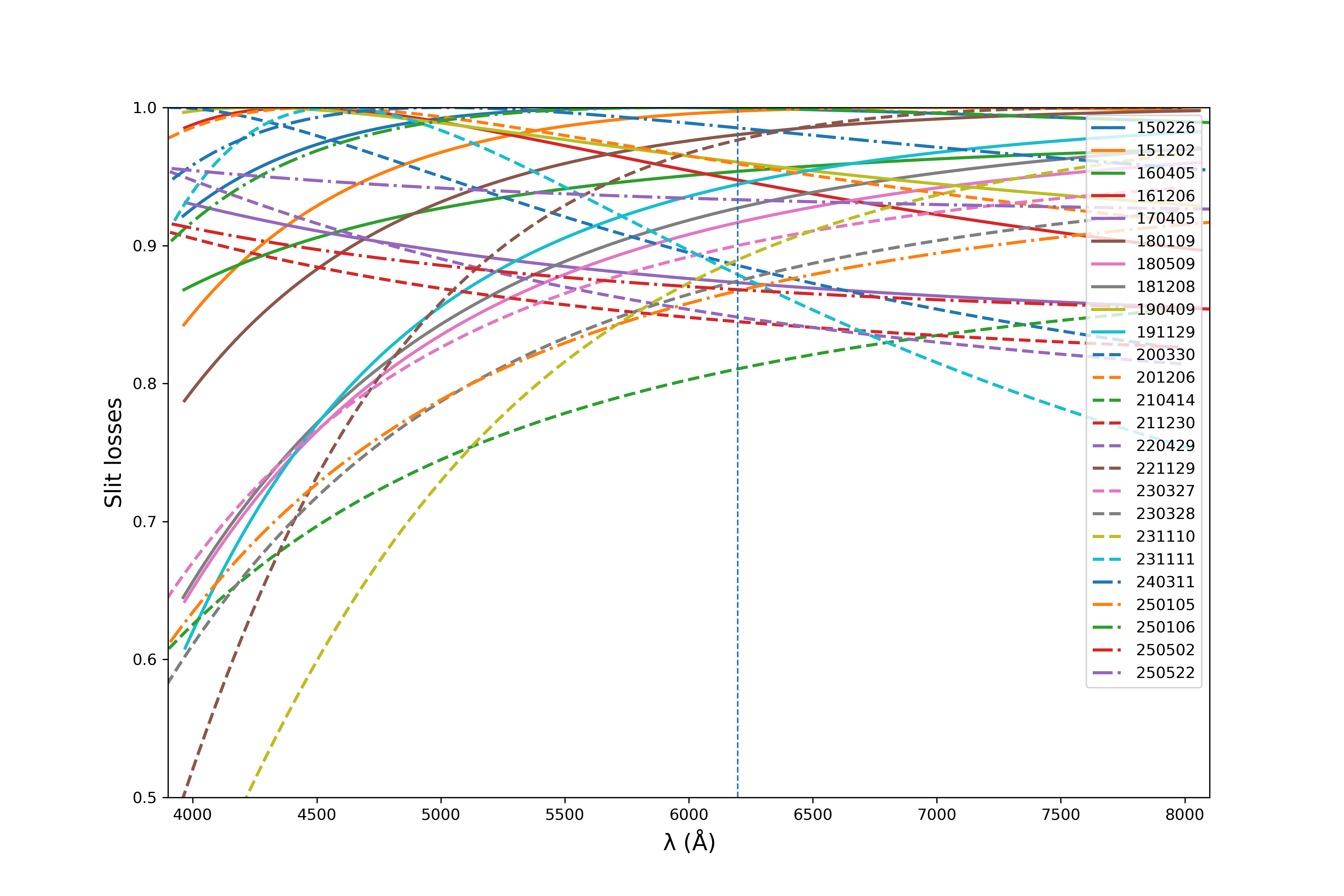}
\caption{Chromatic slit losses for each observation night. Slit losses on a given 
night are estimated using the offsets for that night (see 
Figure~\ref{fig:offset}) and a circular Gaussian source with an observationally 
motivated width (see main text).}
\label{fig:slitloss}
\end{figure} 

An additional problem is the differential atmospheric refraction. At each 
observation epoch, the slit was oriented along the line joining the two quasar 
images (see Sect.~\ref{sec:monitor}), so its position angle did not coincide with 
the parallactic angle. This deviation from the parallactic angle caused DAR, which 
shifted both images across the slit and increased the light (slit) losses, 
especially at blue wavelengths \citep{1982PASP...94..715F}. The LT is not equipped 
with a DAR corrector, and thus we developed a procedure to numerically correct 
these wavelength-dependent slit losses leading to spectral artefacts. Our 
procedure relied on the analysis by \citet{1982PASP...94..715F}, who provided a 
method to derive DAR-induced offsets in function of the deviation from the 
parallactic angle, the airmass value, and the weather conditions. 

The DAR offsets were calculated relative to the reference wavelength (6200 \AA), 
and results at all epochs are shown in Figure~\ref{fig:offset}. For this reference 
wavelength, the quasar images are not on the slit axis, but shifted across the 
slit by the $\delta y$ offsets inferred from the through-slit acquisition frames 
(see above). Additionally, we estimated the chromatic slit losses at all epochs 
for a circular Gaussian source with width $\sigma$, using $\sigma$ values obtained 
by analysing the through-slit acquisition frames (see Figure~\ref{fig:slitloss}).
After correcting for instrument responses and slit losses, we also used IO:O 
$r$-band fluxes to properly calibrate the quasar spectra.

\section{Auxiliary spectra} 
\label{sec:appendix2}

In addition to the set of 25 LT/SPRAT spectra, we considered spectroscopic 
observations from other facilities at four different epochs (vertical dashed lines 
in Figure~\ref{fig:lcs}). Although these data can only yield simultaneous spectra 
of A and B at each observation epoch, and therefore the B/A spectral flux ratio 
could be affected by intrinsic variations, the observations were done using 8$-$10 
m class telescopes and complement our LT monitoring programme.  

\begin{figure}[h!]
\centering
\includegraphics[width=9cm]{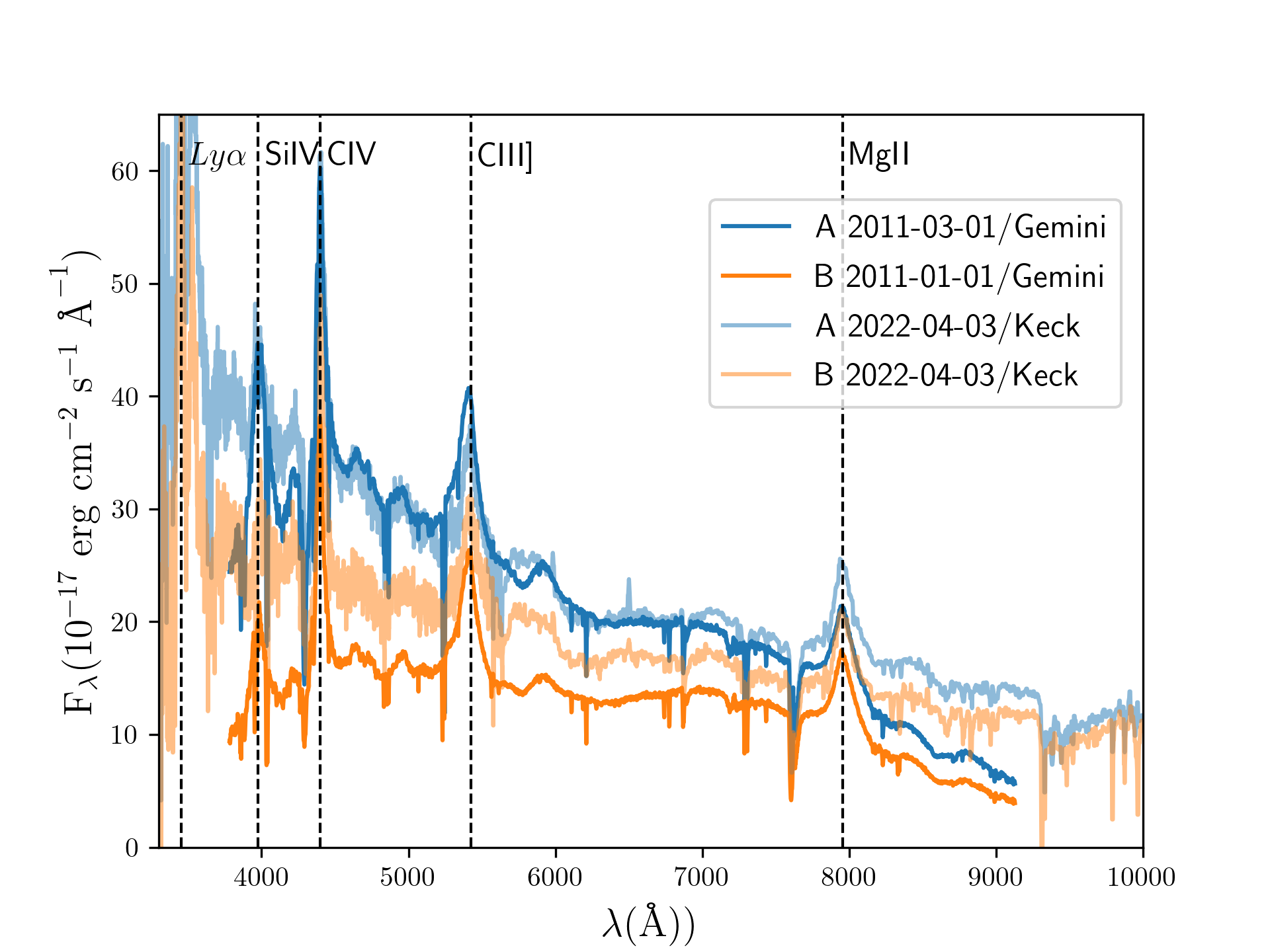}
\caption{Auxiliary spectra of \object{SDSS J1001+5027} from near-UV to near-IR.}
\label{fig:aux}
\end{figure}

We downloaded and processed archive data of the Gemini North 
Telescope\footnote{https://archive.gemini.edu/searchform} (GMOS instrument, 1 
March 2011) and the Keck 
Telescope\footnote{https://koa.ipac.caltech.edu/cgi-bin/KOA/nph-KOAlogin} (LRIS 
instrument, 3 April 2022). These spectroscopic data allowed us to extract spectra 
for the two quasar images extending up to 9000$-$10\,000 \AA\ (see 
Figure~\ref{fig:aux}). We also used Subaru Telescope/HDS spectra on 27 January 
2016 and 19 January 2017 that were obtained by \citet{2018ApJ...854...69M} and 
cover the spectral region around the \ion{C}{iv} emission line.

\end{appendix}

\end{document}